\documentclass[11pt]{article}
\usepackage{amssymb,amsmath}
\usepackage{latexsym}

\newtheorem{ex}{Example}[section]

\def\buildrel#1_#2^#3{\mathrel{\mathop{\kern 0pt#1}\limits_{#2}^{#3}}}

\newcommand{\p}{{\partial}{}}

\newcommand{\re}[1]{(\ref{#1})}

\newcommand{\mysection}[1]{\section{#1}\setcounter{equation}{0}}
\begin{document}

\begin{flushright}
\parbox{5cm}{
\begin{flushleft}
NSF-KITP-09-96 \\ 
ESI - 2149
\end{flushleft}
}
\end{flushright}


\begin{center}
{\Large \bf 
Boundary conditions for spacelike and timelike warped $AdS_3$ \vspace{5pt} \\ 
spaces in topologically massive gravity 
}
\vspace*{0.5cm}\\
{Geoffrey Comp\`ere$^{\flat}$ and
St\'ephane Detournay$^{\natural}$, 
}
\end{center}
\vspace*{0.1cm}
\begin{center}
$^{\flat}${\it Department of Physics, 
University of California, Santa Barbara, \\ 
Santa Barbara, CA 93106, USA
} 
\vspace*{0.5cm}\\
$^{\natural}${\it Kavli Institute for Theoretical Physics, \\ 
University of California, Santa Barbara, CA 93106, USA
} 
\vspace{0.5cm}\\
{\tt gcompere@physics.ucsb.edu}  \\ 
{\tt detourn@kitp.ucsb.edu} \qquad 
\end{center}

\vspace{0.5cm}

\begin{abstract}
We propose a set of consistent boundary conditions containing the spacelike warped black holes solutions of Topologically Massive Gravity. We prove that the corresponding asymptotic charges whose algebra consists in a Virasoro algebra and a current algebra are finite, integrable and conserved. A similar analysis is performed for the timelike warped $AdS_3$ spaces which contain a family of regular solitons. The energy of the boundary Virasoro excitations is positive while the current algebra leads to negative (for the spacelike warped case) and positive (for the timelike warped case) energy boundary excitations. We discuss the relationship with the Brown-Henneaux boundary conditions. 

\vspace{12pt}
Pacs: 04.20.-q,04.60.-m,04.70.-s,11.30.-j
\end{abstract}

Topologically Massive Gravity \cite{Deser:1981wh,Deser:1982vy}  (TMG) has recently received a great deal of attention since the conjecture by Li, Song and Strominger \cite{Li:2008dq}  that the theory is chiral at a particular point in parameter space. Although counter-arguments to the original proposal appeared (regarding chirality and unitarity) \cite{Carlip:2008jk, Grumiller:2008qz, Grumiller:2008es, Giribet:2008bw}, it was shown that a refined version of the conjecture remains intact upon truncating the non-chiral degree of freedom, which in turn could lead to a consistent quantum theory \cite{Maloney:2009ck}.
It has also been known since the work of \cite{Deser:1981wh,Deser:1982vy} that $AdS_3$ space is an unstable background solution of TMG with Lagrangian
\begin{equation}\label{TMGL}
 I_{TMG} = \frac{1}{16 \pi G} \left[\int_M  d^3x \, \sqrt{-g} (R+ \frac 2 {l^2}) + \frac{1}{\mu} \; I_{CS} \right]
\end{equation}
away from the chiral point $\mu l= 1$. The gravitational Chern-Simons term $I_{CS}$ in \re{TMGL} is given by
\begin{equation} 
 I_{CS} = \frac{1}{2} \int_M d^3x \, \sqrt{-g} \varepsilon^{\lambda \mu \nu} \Gamma^\alpha_{\lambda \sigma} \left(\partial_\mu \Gamma^\sigma_{\alpha \nu} + \frac{2}{3} \Gamma^{\sigma}_{\mu \tau}\Gamma^{\tau}_{\nu \alpha}\right).
\end{equation}
We choose $\mu >0$ without loss of generality and we will set $G=1$. The equations of motion of TMG are given by
\begin{equation}\label{EOM}
E_{\mu\nu} \equiv G_{\mu \nu} - \frac{1}{l^2}g_{\mu\nu}+\frac{1}{\mu} C_{\mu\nu}=0
\end{equation}
where $C_{\mu\nu}$ is the Cotton tensor.
 
It has been proposed in \cite{Anninos:2008fx} that another background, the so-called {\itshape spacelike warped $AdS_3$}  space ($WAdS_3$) could actually constitute a stable vacuum of the theory (this question has been investigated in the recent paper \cite{Anninos:2009zi}). Although less symmetric than its unwarped cousin, it shares many of its features. In particular, for 
\begin{eqnarray}
\nu \equiv \frac{\mu l }{3} > 1,
\end{eqnarray}
there exist regular black hole solutions \cite{Moussa:2003fc} that can be obtained by performing discrete identifications in $WAdS_3$, much like BTZ black holes are obtained from $AdS_3$. 

One interest of these warped geometries is that TMG with ``warped boundary conditions'' has been conjectured to be dual to a two-dimensional CFT with two unequal central charges \cite{Anninos:2008fx}
\begin{eqnarray}
c = \frac{(5 {\nu}^2 + 3)l}{ {\nu}({\nu}^2 + 3)},\qquad \bar c = \frac{4 \nu l }{\nu^2+3}.
\end{eqnarray}
Let us note that such a CFT should display particular properties under parity transformations. In $AdS_3$ space, the two asymptotic Virasoro algebras get mapped into each other under a parity transformation. Indeed, the left and right movers are switched, which implies in particular that the two Virasoro central charges $\frac{3l}{2G}(1\pm \frac{1}{\mu l})$ get switched. The two conjectured Virasoro algebras appearing in warped geometries do not transform similarly under parity. Under $\nu \rightarrow -\nu$, the central charges $c$ and $\bar c$ just change sign, which can be reabsorbed into a redefinition of the Virasoro generators $L_{n} \rightarrow - L_{-n}$ and $\bar L_{n} \rightarrow - \bar L_{-n}$. Therefore, contrary to the $AdS_3$ case, the two sectors of the conjectured CFT should transform separately under parity.

The first central charge $c$ has been recovered from a classical asymptotic symmetry analysis, based on a very restricted phase space consisting of warped black holes and their descendants \cite{Compere:2008cv}. Boundary conditions including warped black holes were written down in \cite{Compere:2007in} but the analysis was done in a different theory, namely Einstein-Maxwell-Chern-Simons theory, and only in the regime of parameters with closed timelike curves. Since the definition of charges and the content of the phase space are determined by the theory considered, it is not obvious that the boundary conditions written in \cite{Compere:2007in} are valid also in TMG. The first purpose of this note is thus to fill this gap and provide a general set of boundary conditions encompassing the spacelike warped black hole solutions in TMG. These boundary conditions could be used as a first step towards proving a positivity theorem for warped geometries.

In the second part of this note, we will extend our analysis to the timelike squashed warped geometries and define boundary conditions including the background timelike warped $AdS$. We will describe a two parameter family of solitons, i.e. solutions of TMG obeying these boundary conditions which are completely regular everywhere, without horizons nor closed timelike curves. As far as we are aware, these solitons are a new class of solutions of TMG which, for example, were not discussed in \cite{Anninos:2008fx} since they do not contain horizons, and which evade the analysis of \cite{Banados:2005da} since they are not solutions of Einstein-Maxwell-Chern-Simons theory.

We conclude by comparing the BTZ metrics and the Brown-Henneaux boundary conditions with the $\nu^2 \rightarrow 1$ limit of spacelike and timelike warped spaces. We will show how the Virasoro algebra appearing in the asymptotics of the warped spaces can be mapped on either of the two $AdS_3$ Virasoro algebras. 

\mysection{Spacelike warped boundary conditions}

We impose the following boundary conditions for $\nu^2 > 1$:
\begin{eqnarray}
g_{TT} &=& 1 + O(R^{-1}), \qquad g_{TR} = O(R^{-2}), \qquad g_{T\Phi} = -2 \frac{\nu}{l} R + O(R^0) ,\nonumber \\
g_{RR}&=& \frac{l^2}{3+\nu^2}\frac{1}{R^2}+O(R^{-3}) , \qquad g_{R\Phi} = O(R^{-1}),\qquad g_{\Phi\Phi} = \frac{3(\nu^2-1)}{l^2}R^2 + O(R),\label{BC}
\end{eqnarray}
for spacetimes admitting a limit $R \rightarrow \infty$. Here $T$ spans the real line and $\Phi \in[0,2\pi]$.
We will moreover impose for the simplicity of the analysis that $g_{TT}$ admits a polynomial expansion as $g_{TT} = 1 + g_{TT}^1 R^{-1}+ g_{TT}^2 R^{-2}+o(R^{-2})$ and similarly for the other components $g_{\mu\nu}$.  We denote by $g_{\mu\nu}^1$ and $g_{\mu\nu}^2$ the first two arbitrary functions appearing in the polynomial expansion of $g_{\mu\nu}$\footnote{In view of the linear analysis of \cite{Anninos:2009zi}, it might be interesting to try to find a more general set of boundary conditions whose coefficients are not only polynomials in $R$ but rational functions of $R$, in the same spirit as the log mode \cite{Grumiller:2008qz} can be incorporated into a larger set of boundary conditions \cite{Grumiller:2008es} than the 
Brown-Henneaux ones \cite{Brown:1986nw} for asymptotically $AdS_3$ spacetimes in TMG at $\mu\, l =1$.}.

We will show that these boundary conditions have to be supplemented by a constraint relating the components $g^1$ and their derivatives that we will introduce and justify at point (v), see equation \eqref{constr1}. We will discuss one additional constraint in point (ix), see \eqref{constr2} that could be enforced if one insists on the positivity of the Virasoro zero mode spectrum. A summary of the boundary conditions including these additional restrictions is presented at point (x). 

These boundary conditions enjoy the following properties:
\begin{description}
\item[(i) Black holes are included in the phase space]

The warped black hole solutions of TMG are given by
\begin{eqnarray}\label{WBH}
ds^2 = dT^2  +(\frac{3}{l^2}(\nu^2-1)R^2 - \frac{4 \mathfrak j l}{\nu}+12 \mathfrak m R )d\Phi^2 
- 4  \frac{\nu}{l} R dT d\Phi +\frac{dR^2}{\frac{3+\nu^2}{l^2}R^2 - 12 \mathfrak m R + \frac{4 \mathfrak j \,l}{\nu}},
\end{eqnarray}
where $\mathfrak m$ and $\mathfrak j$ are two parameters. These solutions obey the fall-off conditions \eqref{BC}. The spacelike warped background geometry, i.e. the geometry \eqref{WBH} for $\mathfrak m = \pm 1/6$, $ \mathfrak j =0$ and $\Phi \in \mathbb R$, can be argued not to be included in these boundary conditions, generalizing the discussion of \cite{Anninos:2009zi}, sect. 2.2. Indeed, this spacetime does not admit any Killing vector with closed orbits, in particular, in any asymptotic region of that spacetime while in \eqref{BC} $\frac{\partial}{\partial \Phi}$ is a Killing vector with closed orbits in the asymptotic region $R \rightarrow \infty$. 

\item[(ii) The phase space is invariant under a Virasoro and a current algebra]

The set of infinitesimal diffeomorphisms
\begin{eqnarray}
l_n &=& (N e^{i n \Phi}+O(R^{-1})  \p_T +(- i n R e^{i n \Phi} +O(R^{0}))\p_R + (e^{i n \Phi} +O(R^{-2}) )\p_\Phi \label{VirBC} ,\\
t_n &=& (N^\prime e^{i n \Phi}+O(R^{-1}))\p_T ,\label{tn}
\end{eqnarray}
where $N$, $N^\prime$ are arbitrary normalization constants leave the boundary conditions invariant. Indeed, one can check that the Lie derivative of the metric with respect to $t_n$ and $l_n$ gives a perturbation obeying the fall-off conditions \eqref{BC}. These generators admit the following commutators 
\begin{eqnarray}\label{ASA}
i[l_m,l_n] &=& (m-n)l_{m+n}, \qquad i[l_m,t_n] = -nt_{m+n}, \qquad [t_m,t_n] =0,
\end{eqnarray}
isomorphic to a semi-direct sum of a Virasoro algebra and a current algebra. 

\item[(iii) The charges are finite]

The equations of motion admit the following expansion 
\begin{eqnarray}
E_{TT}& =&R^{-2} E^2_{TT} + O(R^{-3}), \qquad E_{TR} = R^{-2} E_{TR}^1+R^{-3} E_{TR}^2 + O(R^{-4}),\nonumber\\ 
E_{T\Phi} &=& E_{T\Phi}^1+R^{-1}E_{T\Phi}^2 + O(R^{-2})\label{EOMR}\qquad 
E_{RR} = R^{-1} E_{RR}^1 +R^{-2} E_{RR}^2+O(R^{-3}), \\
E_{R\Phi} &=& R E_{R\Phi}^1+E_{R\Phi}^2  + O(R^{-1}), \qquad E_{\Phi\Phi} = R^{-1}E_{\Phi\Phi}^1+R^{-2}E_{\Phi\Phi}^2 + O(R^{-3}), \nonumber
\end{eqnarray}
where the equations $E_{\mu\nu}^1 = 0$ depend only linearly on $g^1_{\alpha\beta}$ while $E_{\mu\nu}^2 = 0$ depends at most quadratically on $g^1_{\alpha\beta}$ and linearly on $g^2_{\alpha\beta}$. 

Using the methods of \cite{Barnich:2001jy,Barnich:2003xg,Barnich:2007bf}, one can define the charge one-form $k_\xi[\delta g ;g]$ associated with the vector $\xi$, see \cite{Compere:2008cv} for an explicit expression in TMG. The infinitesimal charge differences $\delta \mathcal T_n$, $\delta \mathcal L_n$ between two solutions of the phase space associated with the asymptotic symmetries $t_n$ and $l_n$ are given by
\begin{eqnarray}
\delta \mathcal T_n &\equiv & \int_0^{2\pi}k_{t_n}[\delta g ; g] =  \frac{N' }{16 \pi } \int_0^{2\pi}e^{i n \Phi} \left( \delta F^{lin}[g^1,\p_T g^1]+ \frac{4 l \nu}{3(\nu^2-1)}\delta E^1_{T\Phi} \right) , \label{deltaTn}\\
\delta \mathcal L_n &\equiv & \int_0^{2\pi}k_{l_n}[\delta g ; g] = -\frac{N}{8 \pi }  \int_0^{2\pi}e^{i n \Phi} \delta E^{1}_{T\Phi} R + O(R^0), \label{deltaLn} 
\end{eqnarray}
where $F^{lin}[g^1,\p_t g^1]$ is a linear functional of the metric coefficients and their first derivative at first order in the $R$ expansion given by
\begin{eqnarray}
F^{lin}[g^1,\p_T g^1] = \frac{(\nu^2+3)^2}{3 l^4}g_{RR}^1+\frac{(\nu^2-1)}{l^2} g_{TT}^1 +\frac{2(\nu^2-3)}{3\nu l} g^1_{T\Phi}+\frac{1}{3}g_{\Phi\Phi}^1-\frac{(3+\nu^2)}{3\nu l}\p_T g^1_{R\Phi}.
\end{eqnarray}
The current charges are therefore finite. When considering a perturbation tangent to the phase space of solutions to TMG, we see using the equations of motion that the Virasoro charges are also finite.

\item[(iv) The charges associated to the current algebra are integrable and conserved]

The \\
infinitesimal charges associated to the current algebra are linear functionals of the asymptotic component of the metric, i.e. they are asymptotically linear. It is then trivial to define the generators associated to a given solution by integrating the infinitesimal charges on the phase space as
\begin{eqnarray}
 \mathcal T_m &\approx &   \frac{1}{16 \pi } \int_0^{2\pi} e^{i m \Phi} F^{lin}[g^1,\p_T g^1]. \label{Tmexpl}
\end{eqnarray}
where $\approx$ means that the equality is valid on-shell and where we choose the $N'$ factor in \re{deltaTn} to be 1. For the black holes we get $\mathcal T_n = \delta_{n,0}\mathfrak m$.

The charges $\mathcal T_m$ are finite and so independent of $R$. We now have to show that the quantities $\mathcal T_m$ are conserved, i.e. are $T$-independent. Observing that
\begin{eqnarray}
E_{TR}^1 \sim \p_T F^{lin}[g^1,\p_T g^1],
\end{eqnarray}
is sufficient to show that when the equations of motions are obeyed, the charges $\mathcal T_m$ are indeed conserved.

\item[(v) The Virasoro charges are integrable modulo a constraint]

In order to describe the Virasoro generators, let us denote by 
\begin{eqnarray}
\Phi^1_I \equiv \{ g^1_{\mu\nu}, \p_\alpha g^1_{\mu\nu},\p_{\alpha_1} \p_{\alpha_2} g^1_{\mu\nu}, \dots\}
\end{eqnarray}
the set of metric components at first order and their derivatives up to the $k$-th derivative where $k$ is a fixed large integer. The abstract index $I$ spans both the metric indices and the derivatives indices. One can define similarly $\Phi^2_I$ in terms of $g^2$. The set of fields $\Phi^1_I$, $\Phi^2_I$, \dots is a convenient way of parameterizing the phase space described by the boundary conditions \eqref{BC}.

On-shell, one finds that the Virasoro charges have the form
\begin{eqnarray}
\delta \mathcal L_n = \frac{1}{16\pi }\int_0^{2\pi} e^{i n \Phi} \sum_{I,J} \left( a_{IJ} \Phi_I^1 \delta \Phi_J^1+ b_I(n) \delta \Phi^1_I  + c_I \delta \Phi^2_I \right) + N \delta \mathcal T_n.\label{formLm}
\end{eqnarray}
for some coefficients $a_{IJ}$, $c_I$ depending only on the scale $l$ and the parameter $\nu$ and $b_I(n)$ depending also explicitly on $n$. The first set of terms is a quadratic functional on the phase space and is not at first sight a $\delta$-exact quantity. However, one can show using a Mathematica code that the matrix $a_{IJ}$ is in fact a symmetric matrix $a_{IJ}=a_{JI}$. The quadratic term $a_{IJ} \Phi_I^1 \delta \Phi_J^1$ can thus be immediately integrated to 
\begin{eqnarray}
a_{IJ} \Phi_I^1 \delta \Phi_J^1 = \delta ( \frac{1}{2} a_{IJ} \Phi_I^1 \Phi_J^1).
\end{eqnarray}

The last term in \eqref{formLm} is pretty subtle. If one chooses a constant normalization $N = constant$, one could integrate the charges to get
\begin{eqnarray}
\mathcal L_n(N = constant) = \frac{1}{16\pi }\int_0^{2\pi} e^{i n \Phi} \sum_{I,J} \left( \frac{1}{2} a_{IJ} \Phi_I^1 \Phi_J^1+ b_I(n)  \Phi^1_I + c_I  \Phi^2_I \right) + constant\, \mathcal T_n.
\end{eqnarray}
However, one would find that the spectrum of $\mathcal L_0$ for the black holes is unbounded from below. Rather, it was pointed out in \cite{Compere:2008cv} that if one chooses the following field-dependent normalization
\begin{eqnarray}
N = \frac{12 l \nu}{3+\nu^2} \mathcal T_0,\label{normN}
\end{eqnarray}
the generators $\mathcal L_0$ evaluated on the black holes' phase space  are given by the expression 
\begin{equation}
\mathcal L^{black \;holes}_0 = c (\frac{3}{2}\mathfrak m^2 - \frac{2}{3 c\,} \mathfrak j) \label{chargesBH}
\end{equation}
which is non-negative for all regular black holes since $\mathcal L^{black \;holes}_0 \geq 0$ is equivalent to the condition of having an horizon. Moreover, with the choice \eqref{normN}, the Virasoro generators \eqref{VirBC} reduce to one set of the $AdS_3$ Virasoro generators \eqref{VirBH} in the limit $\nu^2 \rightarrow 1$, as will be shown in section \ref{sec:comparison}.  Let us see what the choice \eqref{normN} implies for a general metric obeying the boundary conditions \eqref{BC}. Note that terms of the form $\mathcal T_0 \delta \mathcal T_n$ would in general not be integrable, not being $\delta-$exact. In order to cure that problem, we restrict our phase space to solutions with fixed $\mathcal T_0$ charges, i.e.
\begin{eqnarray}
 \mathcal T_0 = \mathfrak m \label{constr1}
\end{eqnarray}
where $\mathfrak m$ is a fixed quantity, not necessarily positive. This condition allows us to integrate the term $\mathcal T_0 \delta \mathcal T_n$ as $\delta ( \mathcal T_0 \mathcal T_n +\delta_{n,0}  \mathcal {\bar L}_0 )$ 
where $ \mathcal {\bar L}_0$ 
is a background charge. The condition \eqref{constr1} can consistently be imposed given that in the asymptotic symmetry algebra \re{ASA}  and in its subsequent realization in terms of charges - see (vii) -, $\mathcal T_0$ commutes with all other generators.
The quantity $\mathfrak m$ therefore labels different representations of the asymptotic symmetry algebra, while states with a fixed label $\mathfrak m$ will be characterized by their $\mathcal L_0$-eigenvalue.

As a consequence of the constraint \eqref{constr1}, the charges $\mathcal L_n$ of a solution $g_{\mu\nu}$ are defined as the integral of $\delta \mathcal L_n$ in the phase space between the reference solution \eqref{WBH} with $\mathfrak j=0$ and $\mathfrak m$ fixed and the solution $g_{\mu\nu}$ plus the background charge of the reference solution. In order to be consistent with \eqref{chargesBH}, we choose that background charge to be $-\frac{6 l \nu}{3+\nu^2}\, (\mathcal T_0)^2 \delta_{n,0}$. The Virasoro charges are finally given by 
\begin{eqnarray}
\mathcal L_n = \frac{1}{16\pi }\int_0^{2\pi} e^{i n \Phi} \sum_{I,J} \left( \frac{1}{2} a_{IJ} \Phi_I^1 \Phi_J^1+ b_I(n)  \Phi^1_I + c_I  \Phi^2_I \right) -  \frac{6 l \nu}{3+\nu^2}\, (\mathcal T_0)^2 \delta_{n,0} +  \frac{12 l \nu}{3+\nu^2} (\mathcal T_0 \mathcal T_n). \label{LnInt}
\end{eqnarray}

\item[(vi) The charges are represented by a Poisson bracket]

In Hamiltonian formalism, it has been shown \cite{Brown:1986ed} that the asymptotic symmetry algebra is represented by a Poisson bracket of conserved charges on-shell up to central terms when (1) the charges are defined asymptotically (they are finite); (2) the charges are integrable; (3) the asymptotic symmetries preserve the phase space. In Lagrangian formalism, according to the Theorem 12 of \cite{Barnich:2007bf}, the asymptotic symmetry algebra is represented by a covariant bracket of conserved charges on-shell up to central terms when in addition (4) a technical assumption $\int_{S} \delta E_{\mathcal L}[\delta g,\delta g] = 0$ holds. The term $E_{\mathcal L}$ which is defined e.g. in (5.4) of \cite{Compere:2007az}. This term only depends on the Lagrangian of the theory at hand, and is at the origin of the difference between the symplectic structures (and hence the conserved charges) in the Barnich-Brandt formalism \cite{Barnich:2001jy} and in covariant phase space methods \cite{Iyer:1994ys} (see  (1.21) and (2.9) of \cite{Compere:2007az}), though in most cases it does not contribute to the charges. 

The first three points have been proven earlier, in  (ii)-(iii)-(iv)-(v). The expression for $E_{\mathcal L}$ in TMG can be found in eq. (10) of \cite{Compere:2008cv}, and can be checked to satisfy the required condition (4). More specifically, we get $E_{\mathcal L}[\delta g,\delta g] = O(\frac 1  R)$ for the boundary conditions \re{BC}.

Therefore, the asymptotic symmetry algebra is represented by a covariant bracket of conserved charges up to central terms. The algebra is given by \cite{Compere:2008cv}
\begin{eqnarray}
i \{\mathcal L_m,\mathcal L_n\} &=& (m-n)\mathcal L_{m+n}+ \frac{c}{12}m^3\delta_{m+n,0}, \nonumber \\
 i\{\mathcal L_m,\mathcal T_n\} &= & -n (\mathcal T_{m+n} - \mathcal T_0 \delta_{m,-n}),\label{ChargeAlgebra}\\
i\{\mathcal T_m,\mathcal T_n\} &=& -\frac{1}{3\, \bar c} m\delta_{m+n,0},\nonumber
\end{eqnarray}
The generator $\mathcal T_0$ takes the constant value $\mathfrak m$, see \eqref{constr1}, and commutes with the Virasoro generators. As mentioned earlier, in contrast to \cite{Compere:2008cv} all charges appearing in \re{ChargeAlgebra} are computed with respect to the background $\bar g_\mathfrak m$ corresponding to \re{WBH} with $\mathfrak j = 0$ and fixed $\mathfrak m$. In particular, we have that $\mathcal L_n = \int_{\bar g_\mathfrak m}^g \delta \mathcal L_n + N_n$ where $N_n =  \frac{3}{2}c\, \mathfrak m^2 \delta_{n,0}$ can be deduced from \eqref{chargesBH}-\eqref{LnInt}. Therefore, from $\{\mathcal L_m , \mathcal L_n\} \equiv \int_S k_{l_m}[\mathcal L_{l_n}g;g] = \int_{\bar g_\mathfrak m}^g \int_S k_{[l_m,l_n]} [\delta g';g'] +  \int_S k_{l_m}[\mathcal L_{l_n}\bar g_{\mathfrak m};\bar g_{\mathfrak m}]$, one finds that the term linear in $m$ of the last term gets cancelled by the normalization $N_n$ introduced in the definition of $\mathcal L_n $.  The Virasoro central charge $c$ reproduces one sector of the black hole entropy \cite{Anninos:2008fx}. It is an open issue whether there exists another set of boundary conditions that would  admit another Virasoro sector with the conjectured central charge $\bar c = \frac{4 \nu l}{3+\nu^2}$, or if that other sector of the CFT is somehow encoded in the central charge of the current algebra. 

It is important to note that the central extension appearing in the current algebra is negative, and that sign cannot be removed by any other choice of normalization of the generators $t_n$. We will go back to that point in (ix).

\item[(vii) The Virasoro charges are conserved]

A non-trivial consequence of the representation theorem is that the Virasoro charges are conserved on-shell, 
\begin{eqnarray}
  \p_T \mathcal L_n = \{\mathcal L_n,\mathcal T_0\} &=& 0.
\end{eqnarray}
Because the expression for $\mathcal L_n$ \re{LnInt} is rather complicated, it is difficult to check explicitly that property using the asymptotic form of the equations of motion \eqref{EOMR}. Since the representation theorem is quite opaque, let us give some more details showing what is non-trivial in the proof that the Virasoro charges are conserved. 

Using the definition of the Poisson bracket, we have
\begin{eqnarray}
  \p_T \mathcal L_n = \{\mathcal L_n,\mathcal T_0\} &=& \int_S k_{l_n}[\mathcal L_{t_0} g;g] \nonumber\\
     &=& \int_S k_{l_n}[\mathcal L_{t_0} g;g] - \int_S k_{l_n}[\mathcal L_{t_0} {\bar g_{\mathfrak m}};{\bar g_{\mathfrak m}}] + \int_S k_{l_n}[\mathcal L_{t_0} {\bar g_{\mathfrak m}};{\bar g_{\mathfrak m}}] .
\end{eqnarray}
The third term is zero because the reference solution is $T$ independent. This term is also the central term appearing in the Poisson bracket between $\mathcal L_n$ and $\mathcal T_0$ which indeed vanishes in \re{ChargeAlgebra}. We can thus write
\begin{eqnarray}
  \p_T \mathcal L_n &=&
\int_S \int_\gamma \left[ \frac{d}{d g'} k_{l_n}[\mathcal L_{t_0} g';g']\right] dg'  \nonumber\\
          &=& \int_S \int_\gamma k_{[l_n,t_0]} [\delta g';g'] dg' \nonumber\\
          &=& 0.
\end{eqnarray}
In these expressions, $\gamma$ is a path in the phase space of solutions connecting the reference metric ${\bar g_{\mathfrak m}}$ to $g$. Thanks to the integrability condition, the integral in the phase space is independent on the path chosen. The second equality captures the non-trivial part of the representation theorem and depends crucially on the integrabity of the charges. It was first proven in Hamiltonian formalism \cite{Brown:1986ed}, and rederived later in Lagrangian formalism, see Prop 8 in \cite{Barnich:2007bf} or in Prop. 13 of \cite{Compere:2007az}. The third equality follows from the algebra \re{ASA} and the linearity of $k_\xi[\delta g ;g]$ in its argument $\xi$.

\item[(viii) The black holes and their Virasoro descendants have a non-negative $L_0$ eigenvalue]  

If a conformal field theory describes the quantization of the classical phase space \eqref{BC}, we expect that for any given value of $\mathfrak m$, the black holes will be associated with primary states in the quantum theory $| l_0 ;\mathfrak m \rangle$ labeled by their $L_0$ eigenvalues that we denote by $l_0$. Black holes states will form a highest-weight state representation of the Virasoro algebra $L_n$ defined as hermitian operators obtained by canonical quantization of the charges $\mathcal L_n$.

If we act on the state $| l_0 ;\mathfrak m \rangle$ with Virasoro boundary excitations $L_{-n}$ for $n \geq 1$, we will raise the zero-eigenmode $L_0$ by $n$ as a consequence of the algebra. In fact, using the algebra \eqref{ChargeAlgebra} where the brackets are replaced by $-i$ times the commutators of the corresponding operators, one can show that the normalized expectation value of $L_0$ will admit the expansion
\begin{eqnarray}
\frac{\langle l_0, \mathfrak m |e^{p L_{n}} L_0 e^{p L_{-n}}| l_0 ;\mathfrak m \rangle}{\langle l_0, \mathfrak m |e^{p L_{n}}  e^{p L_{-n}}| l_0 ;\mathfrak m \rangle} = \frac{\langle l_0, \mathfrak m |L_0| l_0 ;\mathfrak m \rangle}{\langle l_0, \mathfrak m | l_0 ;\mathfrak m \rangle}(1+ 2 p^2 n^2 (l_0+\frac{c\,}{24}n^2) + O(p^3)).\label{L0CFT}
\end{eqnarray}

Let us now show that one can obtain exactly that expectation value from the classical charge analysis, which will provide a consistency check for the existence of a CFT. Classically, acting on the black holes with boundary excitations (beyond the linear level) consists in performing finite diffeomorphisms on the black hole metric. The general finite diffeomorphism generated by exponentiating the algebra is given by
\begin{eqnarray}
T^{new} &=& T + T(\Phi ;p)+ N (L(\Phi ;p) - \Phi), \nonumber \\
R^{new} &=& R / L^\prime(\Phi ; p),\label{eqdiffeo}\\
\Phi^{new} &=&  \nonumber L(\Phi ; p)
\end{eqnarray}
where $L(\Phi ;p)$, $T(\Phi ;p)$ are $\Phi-$periodic functions - with $L(\Phi ;p)$ single-valued on the circle $[0,2\pi]$ and $T(\Phi ;p)$ single-valued on $\mathbb R$ -  which reduce to $L(\Phi ;p=0) = \Phi$, $T(\Phi ;p=0)= 0$ when the diffeomorphism parameterized by $p$ is the identity and the prime denotes the derivative with respect to $\Phi$. For example, the finite real diffeomorphism associated with the real generator $l_n + l_{-n}$ obeys 
\begin{equation}
L(\Phi ;p) = \Phi + 2 p \cos{(n \Phi)} -n \sin{(2 n \Phi)}p^2 + O(p^3)
\end{equation}
The conserved charges associated to the transformed black hole metric can be computed using the methods of \cite{Barnich:2001jy,Compere:2007az}. After an integration by parts in $\Phi$, we get the result
\begin{eqnarray}
\mathcal T_m &=& - \frac{1}{6 \pi \bar c\,} \int_0^{2 \pi} d\Phi e^{i m \Phi} T'(\Phi ;p) +\delta_{m,0} \mathfrak m,\\
\mathcal L_m &=& \frac{1}{2 \pi} \int_0^{2 \pi} d\Phi e^{i m \Phi} (L'(\Phi ;p))^2 \;\mathcal L^{primary}_0 - \frac{1}{12 \pi \bar c\,} \int_0^{2 \pi} d\Phi e^{i m \Phi} (T'(\Phi ;p))^2 \nonumber\\
&+& \frac{c\,}{24} \frac{1}{2 \pi}   \int_0^{2 \pi} d\Phi e^{i m \Phi} \left[ (\frac{L''(\Phi ;p)}{L'(\Phi ;p)})^2 +2 i m \frac{L''(\Phi ;p)}{L'(\Phi ;p)} \right].
\end{eqnarray}
In particular, when acting only with Virasoro generators on the geometry, the eigenvalue $\mathcal L_0$,
\begin{eqnarray}
\mathcal L_0 &=& \frac{1}{2 \pi} \int_0^{2 \pi} d\Phi (L'(\Phi ;p))^2 \;\mathcal L^{primary}_0 + \frac{c\,}{24} \frac{1}{2 \pi}   \int_0^{2 \pi} d\Phi  (\frac{L''(\Phi ;p)}{L'(\Phi ;p)})^2 
\end{eqnarray}
is manifestly non-negative. Notice that in the linear theory around the black holes, i.e. at linear order in $p$, the energy $\mathcal L_0$ is unaffected by the addition of a boundary excitation. At the next-to linear order in perturbation theory around the black holes, the energy
\begin{eqnarray}
\mathcal L_0 &=& \mathcal L^{primary}_0 + p^2 \left( 2 n^2 (\mathcal L^{primary}_0 +\frac{c}{24} n^2) \right) + O(p^3).
\end{eqnarray}
gets shifted by $\frac{c}{24} n^2$ and rescaled by a power of $2n^2$ which exactly reproduces the expectations \eqref{L0CFT} from the dual CFT.

\item[(ix) The current algebra lowers the $\mathcal L_0$ eigenvalue] 

In the last point, we got the expression for $\mathcal L_0$ upon acting with a general diffeomorphism on the black hole metric. In particular, upon acting with the current algebra only, one lowers the $\mathcal L_0$ eigenvalue as
\begin{eqnarray}
\mathcal L_0 &=& \mathcal L^{primary}_0 - \frac{1}{12 \pi \bar c} \int_0^{2 \pi} d\Phi  (T'(\Phi ;p))^2. \label{chargeL0lowered}
\end{eqnarray}
The classical geometries corresponding to these states can be written in compact form as 
\begin{eqnarray}
ds^2 &=& ds^2_{WBH} + 2 T^\prime(\Phi ;p) dT d \Phi + \left( -\frac{4\nu }{l}T^\prime(\Phi ;p) R + (T^\prime(\Phi ;p))^2  \right) d\Phi^2,\label{ChargeL0}
\end{eqnarray}
where $ds^2_{WBH}$ is the black hole metric \eqref{WBH}. One can in fact reproduce the result \eqref{chargeL0lowered} 
from an argument  similar to \re{L0CFT} by assuming that the black hole states $| l_0 ;\mathfrak m \rangle$ also form a highest weight state representation of the current algebra\footnote{The classical expectation value \eqref{chargeL0lowered} is also reproduced if one assumes that the black holes states $| l_0 ;\mathfrak m \rangle$ form a \emph{lowest} weight state representation of the current algebra. Indeed, the classical result is only sensitive to the real sum $t_n \pm t_{-n}$ and cannot distinguish between positive and negative modes.}. The negative sign in \eqref{chargeL0lowered} is a consequence of the algebra \eqref{ChargeAlgebra} as one can easily check.

Since the resulting $\mathcal L_0$ is unbounded from below, it might be desirable to restrict the phase space to remove the current algebra from the asymptotic symmetry algebra. Imposing the condition
\begin{eqnarray}
\mathcal T_{m} = \delta_{m,0} \mathfrak m, \label{constr2wrong}
\end{eqnarray}
is inconsistent with the algebra \eqref{ChargeAlgebra} because the current algebra is centrally-extended. One can instead truncate the phase space by imposing the following condition
\begin{eqnarray}
g_{\Phi\Phi}^1 + \frac{3(\nu^2-1)}{\nu l}g_{T\Phi}^1 = 12 \mathfrak m ,\label{constr2}
\end{eqnarray}
which is consistent with the black holes solutions and preserved by the action of the Virasoro diffeomorphisms but \emph{not preserved} by the current algebra. This condition was found by inspection.

\item[(x) Summary of the boundary conditions]

The final asymptotic spacelike warped boundary conditions consist in the fall-off conditions \eqref{BC} together with imposing a fixed $\mathcal T_0$ sector \eqref{constr1} and removing ``by hand'' the current algebra \eqref{constr2} if the operator $\mathcal L_0$ is to be bounded from below:
\begin{equation}
\mathcal T_0 = \mathfrak m \,\,,\qquad g_{\Phi\Phi}^1 + \frac{3(\nu^2-1)}{\nu l}g_{T\Phi}^1 = 12 \mathfrak m ,\label{constraints}
\end{equation}
where $\mathcal T_0$ is the charge associated to the generator $\partial_T$. 

The asymptotic symmetry algebra then only consists in the Virasoro algebra $l_n$ and the generator $t_0$. The generators $\mathcal T_n$ are well-defined and enter part of the definition of the charges $\mathcal L_n$ but these charges do not act on the phase space via a Poisson bracket once the condition \eqref{constr2} is enforced because the vector $t_n$ is no longer tangent to the phase space. 

It is an open question to see if all regular solutions of TMG obeying these boundary conditions have a positive $\mathcal L_0$. In view of the results of  \cite{Anninos:2009zi}, one could conjecture that this is indeed the case. These boundary conditions are not sufficient to explain the black hole entropy of the general class of black holes \eqref{WBH} since a second Virasoro is missing. However, the Cardy formula of the restricted class of black holes for which $\mathcal T_0 = \mathfrak m = 0$ (and $\mathfrak j \leq 0$) can reproduce the black hole entropy using only the Virasoro algebra $\mathcal L_n$, in which case the underlying CFT, if it exists, should be chiral.  It is another open question to find if a second Virasoro algebra could be defined in the generic case.

\end{description}

\mysection{Boundary conditions for timelike warped $AdS_3$}

It turns out that our analysis goes through for timelike warped $AdS_3$ spaces (see \cite{Anninos:2008fx} and references therein) by means of an analytic continuation from the spacelike warped case $T \rightarrow i T$, $\Phi \rightarrow i \Phi$, $R \rightarrow -R$ with some differences that we will emphazise. Let us impose the boundary conditions for $\nu^2 < 1$ :
\begin{eqnarray}
g_{TT} &=& -1 + O(R^{-1}), \qquad g_{TR} = O(R^{-2}), \qquad g_{T\Phi} = -2 \frac{\nu}{l} R + O(R^0) ,\nonumber \\
g_{RR}&=& \frac{l^2}{3+\nu^2}\frac{1}{R^2}+O(R^{-3}) , \qquad g_{R\Phi} = O(R^{-1}),\qquad g_{\Phi\Phi} = \frac{3(1-\nu^2)}{l^2}R^2 + O(R).\label{BCT}
\end{eqnarray}
These boundary conditions enjoy the following properties:
\begin{description}
\item[(i) Solitons and the background timelike warped $AdS_3$ are included in the phase space]

Performing the analytic continuation $T \rightarrow i T$, $\Phi \rightarrow i \Phi$, $R \rightarrow -R$ on the black hole metrics \eqref{WBH}, we get 
\begin{eqnarray}\label{WSOL}
ds^2 = -dT^2  +(\frac{3}{l^2}(1- \nu^2)R^2 + \frac{4 \mathfrak j l}{\nu}+12 \mathfrak m R )d\Phi^2 
- 4  \frac{\nu}{l} R dT d\Phi +\frac{dR^2}{\frac{3+\nu^2}{l^2}R^2 + 12 \mathfrak m R + \frac{4 \mathfrak j \,l}{\nu}}. 
\end{eqnarray}
Here the range of the coordinates is as follows: $T \in \mathbb R$, $R \in \mathbb R$ and $\Phi \in [0,2 \pi]$. The sign of $\mathfrak m$ is unphysical since the solitons with $-\mathfrak m$ are related to those with $+\mathfrak m$ by the change of coordinates $R \rightarrow -R$, $\Phi \rightarrow -\Phi$.

When $\nu^2 > 1$, these metrics describe spacetimes with closed timelike curves and conical defects that were found as solutions of Einstein-Maxwell-Chern Simons theory by performing discrete identification in the three-dimensional G\"odel spacetime in \cite{Banados:2005da, Compere:2007in}. When $\nu^2 < 1$, these pathologies can be avoided. These metrics do not admit closed timelike curves (CTCs) at $R \rightarrow \pm \infty$. The relationship $g^{RR} = \frac{4\nu^2}{l^2}R^2+g_{\Phi\Phi}$ implies that if there are regions of spacetime where there are closed timelike curves - where $g_{\Phi\Phi} < 0$ - these curves are not hidden by an horizon since if one starts in the asymptotic region and lowers $R$, CTCs will be encountered first before reaching the horizon. The only two ways out are (i) $g_{\Phi\Phi}$ and $g^{RR}$  vanish at the same time or (ii) $g_{\Phi\Phi}$ is always positive. The former case  implies $\mathfrak j = 0$. Conical singularities then appear at $R=0$ unless $\mathfrak m = \pm \frac{1}{6}$ or $\mathfrak m = 0$. These special solutions are just the timelike squashed $SL(2,\mathbb R) \times U(1)$ invariant geometry (with $\mathfrak m = \pm \frac{1}{6}$), see e.g. \cite{Anninos:2008fx}, and the zero mass solution $\mathfrak m = \mathfrak j = 0$. In situation (ii), $g_{\Phi\Phi}$ is always positive which requires that the minimal value of $g_{\Phi\Phi}$ denoted by $\mathfrak k^2$, 
\begin{equation}
\mathfrak k^2 =  \frac{4 l }{\nu}\left( \mathfrak j - \frac{3 l \nu}{1-\nu^2}  \mathfrak m^2 \right), \label{condreg}
\end{equation}
be always positive. If the angular momentum $\mathfrak j$ is positive enough such that it obeys \eqref{condreg}, $g_{\Phi\Phi} >0$ and $g^{RR} > 0$ which implies that the solutions \eqref{WSOL} are regular everywhere. 

In summary, the boundary conditions \eqref{BCT} contain regular solitons with the range of parameters $\mathfrak k^2 \geq 0$ (which include the zero mass soliton $\mathfrak m = \mathfrak j = 0$) and the timelike squashed background. They also contain geometries with naked conical singularities and closed timelike curves.

\item[(ii) The phase space is invariant under a Virasoro algebra and a current algebra] 

The \\ same set of infinitesimal diffeomorphisms
\begin{eqnarray}
l_n &=& (N e^{i n \Phi}+O(R^{-1})  \p_T +(- i n R e^{i n \Phi} +O(R^{0}))\p_R + (e^{i n \Phi} +O(R^{-2}) )\p_\Phi \label{VirBC2} \\
t_n &=& (N^\prime e^{i n \Phi}+O(R^{-1}))\p_T \nonumber
\end{eqnarray}
where $N$, $N^\prime$ are arbitrary normalizations constants leave the boundary conditions invariant.

\item[(iii) The  charges are finite]

The proof is similar to the one explained in the spacelike case except for a few irrelevant signs in the intermediate expressions.

\item[(iv)  The charges associated to the current algebra are integrable and conserved]

The\\ proof of integrability and conservation is similar to the spacelike case. Note that for the solitons \eqref{WSOL} we have $\mathcal T_0 = - \mathfrak m$.

\item[(v) The Virasoro charges are integrable modulo a constraint]

Following the spacelike case, one has to impose the constraint that $\mathcal T_0$ be constant on the phase space in order that the Virasoro charges be integrable. We impose
\begin{equation} 
\mathcal T_0 = - \mathfrak m,\label{constr12}
\end{equation}
where $\mathfrak m$ is a constant on the phase space. Once again, different classical phase spaces are labeled by the real number $\mathfrak m$. 

The field-dependent normalization $N$ in \eqref{VirBC2} could be chosen such that the zero mode $\mathcal L_0$ be always non-negative for the solitons obeying \eqref{condreg}. One could choose 
\begin{eqnarray}
N = \frac{8\nu l}{1-\nu^2} \mathfrak m - \beta \mathfrak m \label{normN2}
\end{eqnarray}
for any $\beta > 0$, leading to the values of $\mathcal L_0 = \frac{3+5\nu^2}{24l\nu} \mathfrak k^2 + \frac{1}{2}\beta \mathfrak m^2$. The constant $\beta$ will be fixed in (viii).

\item[(vi) The charges form a representation of the asymptotic symmetry algebra]

The charge algebra 
\begin{eqnarray}
i \{\mathcal L_m,\mathcal L_n\} &=& (m-n)\mathcal L_{m+n}+ \frac{c\,}{12}m^3\delta_{m+n,0}, \nonumber \\
 i\{\mathcal L_m,\mathcal T_n\} &= & -n (\mathcal T_{m+n} - \mathcal T_0 \delta_{m,-n}),\label{ChargeAlgebra2}\\
i\{\mathcal T_m,\mathcal T_n\} &=& \frac{1}{3 \bar c} m\delta_{m+n,0}.\nonumber
\end{eqnarray}
differs from the one obtained in the spacelike case only by the sign of the central charge in the current algebra which is now positive. 

The Virasoro generators are conserved (property (vii)) as a consequence of the representation theorem.

\item[(viii) The solitons and their Virasoro descendants have a non-negative $\mathcal L_0$ eigenvalue] 

The descendants of the solitons are defined by acting with the diffeomorphisms \eqref{eqdiffeo}. It turns out that there is only one value of $\beta$ in \eqref{normN2} such that the $\mathcal L_0$ charge be always non-negative for regular solitons and their Virasoro descendants. Let us choose that value $\beta = \frac{3+5\nu^2}{3+\nu^2}\frac{4\nu l}{1-\nu^2}$. We then get that the normalisation 
\begin{eqnarray}
N = \frac{12 \nu l}{3+\nu^2} \mathfrak m  \label{normN3}
\end{eqnarray}
is identical to the one used for the black holes while the $\mathcal L_0$ charge of the solitons \eqref{WSOL} is given by
\begin{eqnarray}
\mathcal L_0^{Sol} = \frac{3+5\nu^2}{6\nu^2} \left( \mathfrak j - \frac{3 l \nu}{1-\nu^2}  \mathfrak m^2 \right) + \frac{2l\nu(3+5\nu^2)}{(1-\nu^2)(3+\nu^2)}\mathfrak m^2.
\end{eqnarray}
The $\mathcal L_0$ charge of the Virasoro descendants of the solitons is given by 
\begin{eqnarray}
\mathcal L_0 &=& \frac{\mathcal L^{Sol}_0}{2 \pi} \int_0^{2 \pi} d\Phi (L'(\Phi ;p))^2  + \frac{c\,}{24} \frac{1}{2 \pi}   \int_0^{2 \pi} d\Phi  (\frac{L''(\Phi ;p)}{L'(\Phi ;p)})^2
\end{eqnarray}
This expression is indeed non-negative for regular solitons. The energy of the solitons goes to $\infty$ in the limit $\nu^2 \rightarrow 1$. This is consistent with the observation that no solitons are known when $\nu^2 \geq 1$. Finally, for the $SL(2,\mathbb R) \times U(1)$ invariant timelike warped background $\mathfrak m = \frac{1}{6}$, $\mathfrak j = 0$, the energy is
\begin{eqnarray}
\mathcal L_0^{background} = -\frac{c\,}{24},
\end{eqnarray}
which reproduces the mass gap of an extremal CFT where the only primary states which are not the identity have an $\mathcal L_0$ eigenvalue larger than $0$. 

\item[(ix) The current descendants of the solitons have a non-negative $\mathcal L_0$ eigenvalue] 

Upon\\ acting with the full asymptotic symmetry algebra \eqref{VirBC2}, one gets solutions of TMG whose $\mathcal L_0$ charge are 
\begin{eqnarray}
\mathcal L_0 &=& \mathcal L^{Sol}_0 + \frac{1}{12 \pi \bar c} \int_0^{2 \pi} d\Phi  (T'(\Phi ;p))^2 \label{ChargeL02}
\end{eqnarray}
Notice the crucial sign difference between \eqref{ChargeL0} and \eqref{ChargeL02} which is a consequence of the positivity of the central charge in the current algebra \eqref{ChargeAlgebra2}. In this sector, the Virasoro zero mode is always non-negative even upon acting on the solitons with the current algebra.

\item[(x) Summary of the boundary conditions] 

The final boundary conditions for the timelike \\ warped geometries consist in the fall-off conditions \eqref{BCT} together with imposing a fixed $\mathcal T_0$ sector \eqref{constr12}.

The asymptotic symmetry algebra consists in both the Virasoro algebra $l_n$ and the current algebra $t_n$. The charge  $\mathcal T_0$ commutes with the other generators so it can be consistently kept fixed. The Virasoro algebra as well as the current algebra are centrally extended with positive central charges. The phase space contains a class of regular solitons whose $\mathcal L_0$ charge is non-negative. It moreover contains the timelike warped background which has a mass gap of $-c/24$ with respect to the solitons. It might be interesting to find out if bulk excitations of TMG obey these boundary conditions and if they have positive energy by extending the work of \cite{Anninos:2009zi} to the timelike warped case.

\end{description}

\mysection{Comparison to Brown-Henneaux boundary conditions}
\label{sec:comparison}

The boundary conditions 
\begin{eqnarray}
g_{tt} &=& -\frac{r^2}{l^2} + O(r^{0}), \qquad g_{tr} = O(r^{-3}), \qquad g_{t\phi} = O(r^0) ,\nonumber \\
g_{rr}&=& \frac{l^2}{r^2} +O(r^{-4}) , \qquad g_{r\phi} = O(r^{-3}),\qquad g_{\phi\phi} = r^2 + O(r^0).\label{BCH}
\end{eqnarray}
where $\phi \in [0,2\pi]$, $t \in \mathbb R$ have been written down to define a phase space for 3d Einstein gravity with negative cosmological constant \cite{Brown:1986nw}. These conditions have been also used in TMG at the chiral point $\mu l = 1$ ($\nu =1/3$)  to define chiral gravity \cite{Li:2008dq}. 

For a generic value of $\nu^2 > 1$, it is not expected that these boundary conditions have anything to do with the spacelike warped boundary conditions \eqref{BC} or when $\nu^2 < 1$ with the timelike warped boundary conditions \eqref{BCT}. However, in the limit $ \nu^2 \rightarrow 1$, one can try to compare them. 

It has been noticed \cite{Banados:2005da} that the regular warped black holes \eqref{WBH} reduce to the BTZ black holes 
\begin{equation}\label{SolBTZGen}
 ds^2 = -(N^\perp)^2 dt^2 + (N^\perp)^{-2} dr^2 + r^2 (d\phi + N^\phi dt)^2 ,
\end{equation}
with
\begin{equation}
  (N^\perp)^2 =-M + \frac{r^2}{l^2} + \frac{J^2}{4 r^2} \quad , \quad N^\phi = -\frac{J}{2 r^2} \quad ,  
\end{equation}
in a rotating frame when $\nu^2 = 1$. On the other hand, we have mentioned earlier that the regular solitons \eqref{WSOL} disappear in the limit $\nu^2 \rightarrow 1$ because their energy becomes infinite. In order to understand the relationship between the BTZ metric and the spacelike \eqref{WBH} and timelike metrics \eqref{WSOL} in more generality, let us introduce two changes of coordinates, parameterized by a sign $\varepsilon_2$ which are valid respectively when $M l + J \neq  0$ ($\varepsilon_2=+1$) and when $M l - J \neq 0 $ ($\varepsilon_2=-1$). In order to write down the rotating frame for the BTZ metrics, let us introduce another  sign  $\varepsilon= \text{sign}(M l + \varepsilon_2 J)$ which is always positive in the range of non-extremal BTZ black holes ($M l + J > 0$ and $M l - J > 0$), always negative in the range of conical defects $M l + J < 0$ and $M l - J < 0$ and is positive or negative in the other cases. The BTZ metric can then be written in the coordinates $(T,R,\Phi)$ given by 
\begin{eqnarray}
 t = \frac{1}{6 \mathfrak m} T , \quad r = \sqrt{12 \mathfrak m R - 4 \varepsilon j l}, \quad \phi = \varepsilon_2\Phi -\frac{\varepsilon_2}{6 \mathfrak m l}T.\label{BTZGBH}
\end{eqnarray}
as 
\begin{eqnarray}\label{WBHred}
ds^2 = \varepsilon dT^2  +( 12 \mathfrak m R - 4 \varepsilon \mathfrak j l )d\Phi^2 
- 4  \frac{R}{l} dT d\Phi +\frac{dR^2}{\frac{4}{l^2}R^2 - 12 \varepsilon \mathfrak m R + 4 \mathfrak j \,l},
\end{eqnarray}
where the new parameters are $\mathfrak m = \frac{\varepsilon_2}{6}\sqrt{|M+\frac{\varepsilon_2 J}{l}|}$ and $\mathfrak j = \frac{\varepsilon \varepsilon_2}{8}J$. The metric \eqref{WBHred} is the limiting case of the spacelike \eqref{WBH} or timelike \eqref{WSOL} spacetimes when $\nu^2 = 1$. Therefore, all non-extremal BTZ black holes are mapped to the spacelike ($\epsilon = +1$) metrics \eqref{WBHred} using either change of coordinates. The extremal black holes can be mapped also to the spacelike metric using one of the appropriate change of coordinates. These changes of coordinates are not valid for the zero mass black hole. The conical defects and the background anti-de Sitter space are mapped to the timelike  ($\epsilon = -1$) metrics \eqref{WBHred}.

The same changes of coordinates \re{BTZGBH} map the Brown-Henneaux boundary conditions \eqref{BCH} to fall-off conditions comparable to \eqref{BC} when $\nu^2 = 1$ with however some differences that we will emphasize by a mark $*$, 
\begin{eqnarray}
g^{diff\;on\;BH}_{TT} &=^*& O(R^{0}), \qquad g^{diff\;on\;BH}_{TR} = O(R^{-2}), \qquad g^{diff\;on\;BH}_{T\Phi} = -2 \frac{R}{l}  + O(R^0) ,\label{BHCdiff}\\
g^{diff\;on\;BH}_{RR}&=& \frac{l^2}{4}\frac{1}{R^2}+O(R^{-3}) , \qquad g^{diff\;on\;BH}_{R\Phi} =^* O(R^{-2}),\qquad g^{diff\;on\;BH}_{\Phi\Phi} =^* 12 \mathfrak m R+O(R^0).\nonumber 
\end{eqnarray}
The condition on $g_{TT}$ is weaker than $g_{TT} = 1$ in \eqref{BC} or $g_{TT} = -1$ in \eqref{BCT} and clearly, it is not possible to rescale $T$ to get both $g_{TT} = \pm 1$. It seems to indicate that there are spacetimes which obey the boundary conditions of Brown-Henneaux that cannot be mapped to spacetimes obeying the spacelike or the timelike warped boundary conditions even in the limit $\nu^2 \rightarrow 1$. Moreover, given that $g_{R\Phi}$ and $g_{\Phi\Phi}$ in \eqref{BC} are less constrained than in \eqref{BHCdiff}, we suspect that some spacetimes obeying the boundary conditions \eqref{BC} or \eqref{BCT} are not diffeomorphic in the limit $\nu^2 \rightarrow 1$ to spacetimes obeying the Brown-Henneaux boundary conditions.

The Brown-Henneaux boundary conditions admit as asymptotic symmetries the two sets of Virasoro generators
\begin{eqnarray}
\xi^t &=& l (T^+(x^+) + T^-(x^-))+O(r^{-2}),\\
\xi^r &=& - r (\p_+ T^+(x^+) + \p_- T^-(x^-))+O(r^{-1}),\\
\xi^\phi &=&  T^+(x^+) - T^-(x^-)+O(r^{-2}).
\end{eqnarray}
with $x^\pm = \phi \pm t/l$ and with right-moving Virasoro generators $l_n$ associated to $T^+$ and left-moving Virasoro's $\bar l_n$ associated with $T^-$. Under the diffeomorphism \eqref{BTZGBH}, these generators transform as
\begin{eqnarray}
\xi^T &=& 6 \mathfrak m \,l  (T^+(x^+) + T^-(x^-))+O(R^{-1}),\nonumber\\
\xi^R &=& - 2 R (\p_+ T^+(x^+) + \p_- T^-(x^-))+O(R^{0}),\label{VirBH}\\
\xi^\Phi &=& (1+\varepsilon_2) T^+(x^+)+(1-\varepsilon_2) T^-(x^-) +O(R^{-1}).\nonumber
\end{eqnarray}
For the change of coordinates with $\varepsilon_2 = +1$, we find that the right-moving Virasoro generators coincides with the asymptotic generators \eqref{VirBC}-\eqref{VirBC2} with the normalisation \eqref{normN} when $\nu^2 = 1$. Similarly, one can use the change of coordinates with $\varepsilon_2 = -1$ to map the left-moving Virasoro generators to the generators \eqref{VirBC}-\eqref{VirBC2} when $\nu^2 = 1$. Therefore, we conclude that the Virasoro algebra found in the asymptotic symmetry algebra of timelike and spacelike spaces can be mapped in the limit $\nu^2 \rightarrow 1$ to both the left and right-moving Virasoro algebras. 

Finally, note that the second set of Virasoro generators in \eqref{VirBH} cannot be extended to define a consistent second Virasoro algebra in warped spaces $\nu^2 \neq 1$. Indeed, the left-moving generators when $\varepsilon_2 = +1$ and the right-moving generators when $\varepsilon_2 = -1$ expanded in modes have the form
\begin{eqnarray}
\xi^T &=& -3 \varepsilon_2 \mathfrak m \,l  e^{i n \left( \Phi - \frac{T}{3 \varepsilon_2 \mathfrak m l} \right)}+O(R^{-1}),\nonumber\\
\xi^R &=& i n R e^{i n \left( \Phi - \frac{T}{3 \varepsilon_2 \mathfrak m l} \right)}+O(R^{0}),\\
\xi^\Phi &=& O(R^{-1}).\nonumber
\end{eqnarray}
They are not the $\nu^2 \rightarrow 1$ limit of any asymptotic symmetry of \eqref{BC}-\eqref{BCT}. If one tries to use these generators for $\nu^2 \neq 1$, one finds that they are associated with infinite charges.

\section*{Acknowledgements}
We are grateful  to the organizers of the "Workshop on 3d gravity" at the Erwin Schrodinger Institute (April 2009), especially Daniel Grumiller, for setting up an extremely productive and lively meeting, as well as the participants for plenty of very interesting and fruitful discussions. We thank Dionysios Anninos, Mboyo Esole and Monica Guica for letting us know about their results \cite{Anninos:2009zi} prior to publication and most of all for motivating the present work. 
We are very grateful to Steve Carlip, Sophie de Buyl, Gaston Giribet, Tom Hartman, Marc Henneaux, Cristian Martinez, Matt Roberts, Ricardo Troncoso, Andy Strominger and in particular to Dionysios Anninos, Monica Guica and Don Marolf for enlightening discussions on topics dealt with in this note.
The work of S.~D. is funded by the European Commission though the grant PIOF-GA-2008-219950 (Home Institution: Universit\'e Libre de Bruxelles, Service de Physique Th\'eorique et Math\'ematique, Campus de la Plaine, 1050 Brussels, Belgium) and also  supported in part by the National Science Foundation under Grant No. PHY05-51164.
 The work of G.~C. is supported in part by the US National Science Foundation under Grant No. PHY05-55669, and 
by funds from the University of California.


\providecommand{\href}[2]{#2}\begingroup\raggedright\endgroup

\end{document}